# Single-crystal growth and magnetic properties of the metallic molybdate pyrochlore $Sm_2Mo_2O_7$

Surjeet Singh, R. Suryanarayanan, Romuald S. Martin, G. Dhalenne, A. Revcolevschi

Université Paris Sud, Laboratoire de Physico-Chimie de l'Etat Solide, UMR8182, Bât 414, Orsay, France.

We have successfully grown $cm^3$-size single crystals of the metallic-ferromagnet $Sm_2Mo_2O_7$ by the floating-zone method using an infrared-red image furnace. The growth difficulties and the remedies found using a 2-mirror image furnace are discussed. Magnetization studies along the three crystalline axes of the compound are presented and discussed based on our recent proposal of an ordered spin-ice ground state for this compound.

Geometrical frustration in a compound occurs when the spins on its lattice cannot simultaneously minimize all their pair-wise energies because of the incompatibility between the lattice symmetry and the spin-spin interactions. The molybdate pyrochlores $R_2Mo_2O_7$ (R = Nd, Sm, Gd,…, Yb and Y) constitute an interesting series of geometrically frustrated magnets, where the spin-frustration arises from the corner-shared tetrahedral networks of magnetic $R^{3+}$ and $Mo^{4+}$ ions (Fig. 1a). The heavier rare-earths (R = Tb to Yb including Y) members of this series are insulating spin-glasses, in sharp contrast with the lighter rare-earth members (R = Nd, Sm and Gd) which are strongly-correlated ferromagnetic metals [1-3]. In $Gd_2Mo_2O_7$ (GMO), for example, the Mo spins order ferromagnetically near $T_c = 60$ K. At low temperature, the Gd spins couple ferromagnetically to the ordered Mo spins resulting into a simple collinear ferromagnetic structure [4, 5]. In $Nd_2Mo_2O_7$ (NMO), on the other hand, the Nd spins couple antiferromagnetically to the ferromagnetically ordered Mo spins ($T_c = 90$ K) but the resulting magnetic structure at low-temperatures is far from being a simple collinear ferrimagnet. Such a contrasting difference in behavior of Gd and Nd analogues is due to crystal field effects which are absent for $Gd^{3+}$ ion due to its half-filled 4f shell. The $Nd^{3+}$ spins order in a highly non-coplanar arrangement where two spins on an elementary Nd tetrahedron *point-in* whereas the two other *point-out* [6, 7], a spin-arrangement for which Harris et al.[8], who first observed it in the insulating pyrochlore $Ho_2Ti_2O_7$, coined the term "spin-ice" (Fig. 1b) in analogy with the arrangement of protons in water-ice (Fig. 1c) where two protons are closer (covalently bonded) and two farther (hydrogen bonded) from the oxygen-ion (black sphere) at the center. The concurrence of metallic and spin-ice states in the same material, NMO, is unique and is believed to be responsible for its highly unconventional magnetotransport behavior, which is well supported from the fact that the corresponding behavior in GMO is rather conventional [5, 6]. These observations motivated interesting theoretical developments [9], however, further progress in this field is hampered by the unavailability of other metallic spin-ice examples. The pyrochlore $Sm_2Mo_2O_7$ (SMO) being a metallic-ferromagnet [10], it is worth asking: what is the ground state of the Sm spins in this metallic pyrochlore? However, due to several specific difficulties in obtaining high-quality, single crystals of SMO, and because of its large neutron absorption cross-section, the nature of the ground state of Sm spins in this compound has remained unexplored. To the best of our knowledge, $Nd_2Mo_2O_7$ is the only molybdate pyrochlore in which an anisotropic nature of transport and magnetic properties has been reported so far. Concerning now the single crystal growth difficulties, we observed, indeed, that while the titanate pyrochlores are easy to obtain in single-crystalline form [11-13], crystal growth of molybdate pyrochlores presents several specific difficulties: (1) controlling the oxidation state of Mo at +4 during the entire crystal growth procedure , (2) decomposition of the pyrochlore $Sm_2Mo_2O_7$ phase near $T_d \sim 600°C$ [1, 14, 15], (3) high vapor-pressure and hence evaporation at already low temperature of $MoO_2$, and (4) high melting temperature ($T_m$ ~1500 °C to 2000 °C) of most of these molybdate pyrochlores.

Recently, based on our detailed magnetization and thermodynamic studies along different crystallographic orientations, we have identified SMO as another example of metallic spin-ice behavior [16]. Here we report on the successful growth of $cm^3$-size single crystals of the metallic-ferromagnet SMO using the floating-zone (FZ) method. We also discuss the growth difficulties encountered and the remedies found to overcome these difficulties. The anisotropic magnetization behavior along the three crystalline axes is also presented. The remarkable magnetocrystalline anisotropy which develops in $Sm_2Mo_2O_7$, at low temperatures, is discussed.

The crystal growth experiments were carried out by the FZ method using two different image furnaces equipped with: (1) four ellipsoidal mirrors (4-mirror) and (2) two ellipsoidal mirrors (2-mirror). The schematics of the latter are shown in Fig. 2. The FZ method consists of establishing a melt between a polycrystalline rod and a seed, which is then moved up slowly along the feed rod, transforming it into a single-crystal (Fig. 2). This method, when adapted on an infrared image furnace offers several distinct features that can be exploited in the single crystal growth experiments of $Sm_2Mo_2O_7$: (1) high temperatures, up to 2000 °C, can be reached, (2) there is a large temperature gradient along the growth-axis, and (3) growth conditions such as growth atmosphere and growth rate can be controlled easily.

The polycrystalline pyrochlore phase was prepared by the standard solid-state reaction route using high-purity $Sm_2O_3$ (99.99 % purity) and $MoO_2$ (99.9 %) from Alfa Aesar as starting materials. The $Sm_2O_3$ powder was fired at 900 °C for 12 hrs in air and the $MoO_2$ powder was compacted and fired at 700 °C under flowing Ar gas for 3-4 hrs prior to mixing. The intimately mixed powders of the starting oxides were compacted in the form of rods (diameter = 5 mm, length = 60 mm) under a hydrostatic pressure of 2.5 kbar. The rods were heat treated under a purified stream of Ar gas at T = 1280 °C for 16 hrs with one intermediate grinding. The purification of the Ar gas (ARCAL 1) was achieved by passing it through a mesh of Zr-Ti getter at T = 600 °C. The heat treatment was terminated by cooling the samples to room temperature in 30-45 min under the Ar atmosphere to avoid decomposition of the SMO pyrochlore phase into $Sm_2MoO_5$ and $MoO_2$ phases. The first crystal growth attempt was made using a 4-mirror image furnace (maximum power = 1*4 = 4 kW). In all the growth experiments we used a polycrystalline seed cut from one end of the feed-rod. The feed and seed-rods were lowered at 5 mm/hr and rotated in opposite directions at 20 rpm. A strong evaporation of $MoO_2$ was observed both from the floating-zone and a good length of the feed-rod above the floatingzone
(the length MN of the feed rod over which the pyrochlore phase had presumably decomposed, see Fig.2). Furthermore, the evaporated $MoO_2$ began to accumulate on the walls of the quartz enclosure reducing the power transfer fraction from the lamps to the floating-zone below a point where it became impossible to melt the feed rod, even at the maximum available furnace power.

The failure of the first growth procedure, which led us to a two-phase micro-structure consisting of $Sm_2Mo_2O_7$ phase and a molybdenum deficient phase (Fig. 3a), gave us two important indications: (1) it is necessary to have a steeper temperature gradient along the growth axis so as to maximize the length NP of the feed-rod (Fig. 2) which is "colder" than the decomposition temperature $T_d$ (~ 600 °C) of SMO, (2) one has to use a higher growth speed so as to minimize the $MoO_2$ loss from the length MN of the feed-rod (Fig.2). As noted in ref. [17], the temperature from the zone-center along the growth axis falls-off more rapidly in a 2-mirror image furnace than in a 4-mirror furnace. It may be possible to have a steep temperature gradient even in a 4-mirror furnace by adjusting the size and geometry of the lamp-filament and the maximum available furnace power, see for example, ref. [18]. The subsequent crystal growth experiments were, therefore, carried out with a 2-mirror image furnace (NEC model

SC-N15HD), putting a 1 % excess of $MoO_2$. The "best" growth, at an increased growth speed of about 10 mm/hr resulted in a 40 mm long, 5 mm diameter crystal ingot.
Thin slices were cut perpendicular to the length at several locations of the ingot and mirror polished for optical microscopy under polarized light, scanning electron microscopy and Energy Dispersive Analysis of X-rays. Using these techniques, a small portion nearly 6 mm thick, taken at 15 mm from the beginning of the growth was found to be of superior quality. While a section from immediately below the "good portion" showed the presence of two slightly disoriented crystals of the pyrochlore phase, the part above it showed the presence of Mo deficient secondary phase (Sm:Mo   17:13). The quality of the "good portion" is further assessed using a Laue back reflection technique. Using this technique the crystals were cut into [100], [110] and [111] oriented samples (Fig. 4). A lattice parameter of 10.420 Å for the crystals of SMO was found from powder x-ray patterns taken on powder obtained by crushing a small single crystal. This value is in good agreement with the value reported earlier [1].

Magnetization studies as a function of temperature, M(T), were carried out along the [100] and [111] axes (Fig. 4a(inset)). The sharp increase below $T_c$ = 80 K is due to the onset of ferromagnetic ordering of the Mo spins, consistent with previous reports [10]. Upon cooling below $T_c$, magnetization along [100] shows a monotonic increase down to T = 2 K but the magnetization along the [111] axis drops sharply below T* = 15 K. In Fig. 4 the field variation of the isothermal magnetization M(H) is shown along the [100], [110] and [111] axes, at T = 2 K and 25 K. The magnetization behavior is strongly anisotropic at T = 2 K, with an easy-axis behavior along [100], a step-like change near $H_C$ = 15 kOe along [111] and a gradual increase of the [110] magnetization. The magnetization of a polycrystalline sample (Fig. 4c), in contrast, tends to saturate at a value significantly smaller than that of the crystal for any field direction. The step in the [111] magnetization isotherm and the easy(hard)-axis behavior along the [100]([110]) axis is consistent with the spin-ice model [19] and a similar behavior is seen in the canonical spin-ices $Dy_2Ti_2O_7$ [20] and $Ho_2Ti_2O_7$ [21]. In the spin-ice state the [100] field direction is an easy-axis of magnetization (Fig. 1b). The decrease in [111] M(T) below T* (Fig. 4a, inset) can be understood as follows: above T* the Mo spins are free to align along the field direction [111], however, below T*, the easy-axis [100] causes the Mo spins to reorient along this direction, resulting in a decrease of the measured magnetization along [111]. Details are discussed elsewhere [16].
To conclude, by optimizing the growth parameters used in a 2-mirror image furnace we have successfully grown $cm_3$-size, high-quality, single crystals of the ferromagnetic-metal SMO. The specific difficulties of single crystal growth and the procedure to overcome these difficulties are discussed. The magnetization behavior observed along the three crystallographic axes is discussed based on our recent proposal of an ordered spin-ice ground state of Sm spins in SMO [16]. This much awaited second example of the metallic spin-ice should provide a useful test ground for the Berry phase theory of anomalous Hall effect first presented in the context of metallic-spin ice $Nd_2Mo_2O_7$ [9].

**Acknowledgements**
We would like to thank Jacques Berthon and Patrick Berthet for many useful discussions. This work is partially funded by the Centre Franco-Indien pour la Promotion de la Recherche Avancée (CEFIPRA) under project no. 3108-1.
**FIGURE CAPTIONS:**
**Fig. 1 (a)** Cubic unit cell of the pyrochlore $R_2M_2O_7$. **(b)** Spin-ice state on an elementary tetrahedron ($T_h$): the spins at the vertices of $T_h$ are confined to the local [111]-axes in a "2-in" and "2-out" arrangement. Arrow at the centre is net moment per $T_h$. **(c)** Arrangement of protons in water-ice.
**Fig. 2 (a)** Schematics of the crystal growth apparatus: **(1, 2)** ellipsoidal reflecting surfaces; **(3, 4)** Halogen lamps; **(5)** infrared radiations; **(6)** growing crystal; **(7)** floating-zone; **(8)** feed rod; **(9)** quartz enclosure. (Inset) typical temperature profile in an image furnace along its growth axis. **(b)**, **(c)** and **(d)** Single crystal specimens of $Sm_2Mo_2O_7$ along different crystal orientations obtained from the as-grown crystal ingot in **(e).**

**Fig. 3** SEM images of crystals grown using: **(a)** 4-mirror-type furnace. The $Sm_2Mo_2O_7$ pyrochlore phase is represented by the grey and the Mo deficient phase (Sm:Mo  17:13) by the white regions, **(b)** 2-mirror type image furnace showing a single phase $Sm_2Mo_2O_7$ phase (Sm:Mo  16.9:15.8).

**Fig. 4** Isothermal magnetization at T = 2, 25 K and Laue back-reflection photographs of $Sm_2Mo_2O_7$ crystal: **(a)** [100]-axis; **(b)** [110]-axis and **(c)** [111]-axis. The isothermal magnetization of a polycrystalline sample of $Sm_2Mo_2O_7$ is shown in (c). Inset in (a) shows the temperature variation of the magnetization of $Sm_2Mo_2O_7$ along the [100] and [111]-axis.

FIG. 1 (singh et al.)

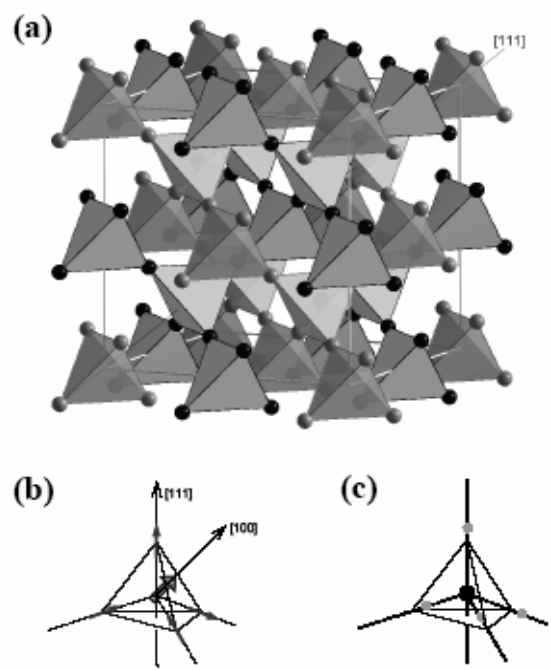

Fig.2

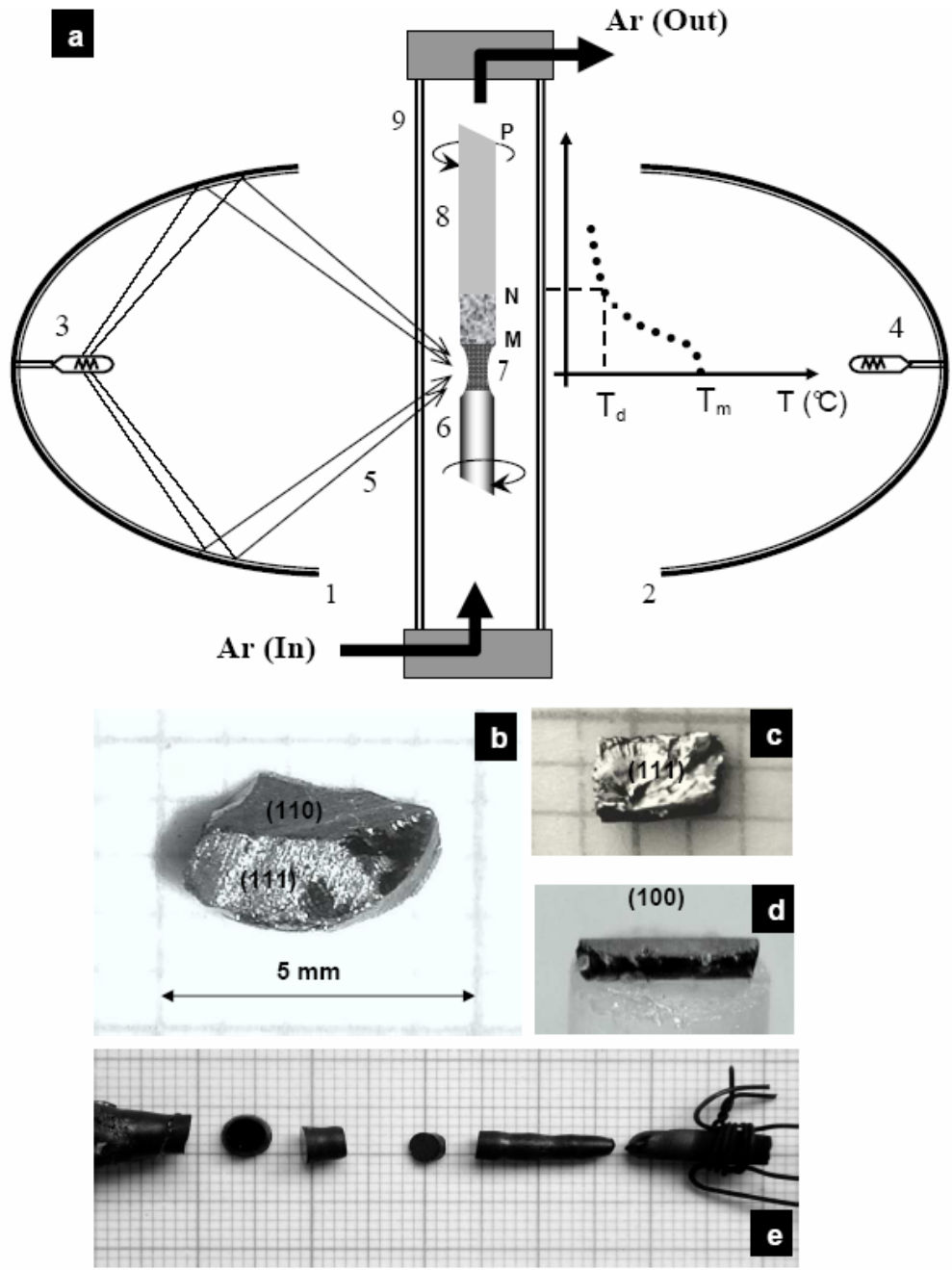

FIG. 3 (singh et al.)

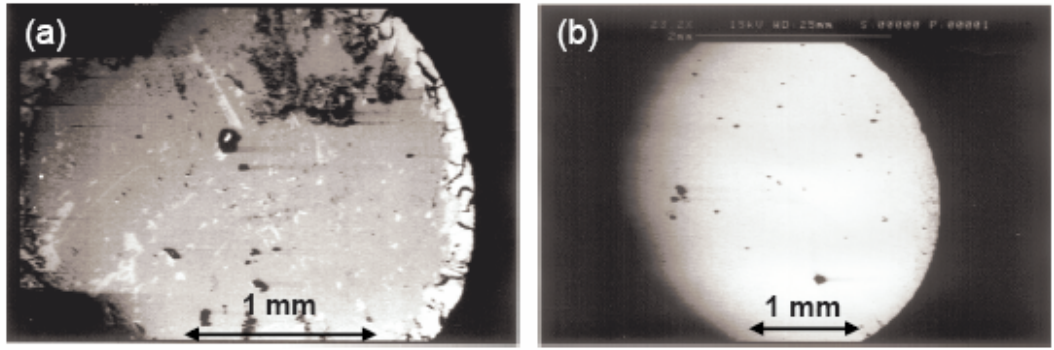

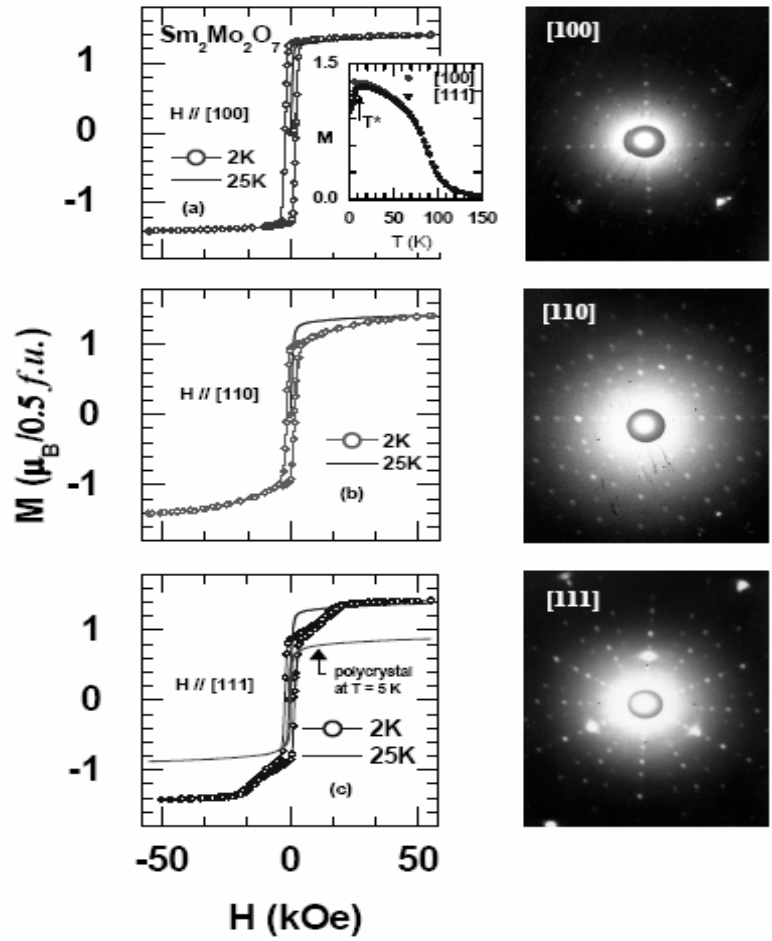